\documentclass[sigconf]{acmart}
\newcommand{\sys}[1]{\textit{Nooks}}
\newcommand{\ed}[1]{{\color{black}
{#1}}}

\AtBeginDocument{%
  \providecommand\BibTeX{{%
    \normalfont B\kern-0.5em{\scshape i\kern-0.25em b}\kern-0.8em\TeX}}}

\copyrightyear{2023}
\acmYear{2023}
\setcopyright{rightsretained}
\acmConference[CHI '23]{Proceedings of the 2023 CHI Conference on Human Factors in Computing Systems}{April 23--28, 2023}{Hamburg, Germany}
\acmBooktitle{Proceedings of the 2023 CHI Conference on Human Factors in Computing Systems (CHI '23), April 23--28, 2023, Hamburg, Germany}\acmDOI{10.1145/3544548.3580796}
\acmISBN{978-1-4503-9421-5/23/04}

\begin{document}

\title[\sys{}: Social Spaces to Lower Hesitations in Interacting with New People at Work]{\sys{}: Social Spaces to Lower Hesitations in Interacting with \\ New People at Work}












\author{Shreya Bali}
\affiliation{%
  \institution{Carnegie Mellon University}
  \city{Pittsburgh}
  \state{Pennsylvania}
  \country{USA}}
\email{sbali@andrew.cmu.edu}

\author{Pranav Khadpe}
\affiliation{%
  \institution{Carnegie Mellon University}
  \city{Pittsburgh}
  \state{Pennsylvania}
  \country{USA}}
\email{pkhadpe@cs.cmu.edu}

\author{Geoff Kaufman}
\affiliation{%
  \institution{Carnegie Mellon University}
  \city{Pittsburgh}
  \state{Pennsylvania}
  \country{USA}}
\email{gfk@cs.cmu.edu}

\author{Chinmay Kulkarni}
\affiliation{%
  \institution{Emory University \& Carnegie Mellon University}
  \city{Atlanta}
  \state{Georgia}
  \country{USA}}
\email{chinmay.kulkarni@emory.edu}

\renewcommand{\shortauthors}{Shreya Bali, et al.}

\begin{abstract}

Initiating conversations with new people at work is often intimidating because of uncertainty about their interests. People worry others may reject their attempts to initiate conversation or that others may not enjoy the conversation. We introduce a new system, \sys{}, built on Slack, that reduces fear of social evaluation by enabling individuals to initiate any conversation as a nook—a conversation room that identifies its topic, but not its creator. Automatically convening others interested in the nook, \sys{} further reduces fears of social evaluation by guaranteeing individuals in advance that others they are about to interact with are interested in the conversation. In a multi-month deployment with participants in a summer research program, \sys{} provided participants with non-threatening and inclusive interaction opportunities, and ambient awareness, leading to new interactions online and offline. Our results demonstrate how intentionally designed social spaces can reduce fears of social evaluation and catalyze new workplace connections.
\end{abstract}




\maketitle

\section{Introduction}
Having conversations with new people is fundamental to how individuals ease into a new workplace~\cite{boothby2018liking} and the resulting social relationships can have far-reaching consequences for both individuals and the organization~\cite{tschan2004work, baron1996interpersonal}. Social relationships at work contribute to well-being and performance by being a vehicle for social support~\cite{kahn1992stress, elfering2002supportive, tschan2004work} and knowledge~\cite{dutton2003power, argote2000knowledge}. At the same time, social relationships promote coordination and organizational citizenship behaviors~\cite{organ1997organizational}, forming the communal underpinnings of organizational effectiveness. However, even if individuals share a mutual desire to interact and develop familiarity, initiating interactions with new people is rarely easy~\cite{boothby2018liking, mitchell2019facilitating, sandstrom2021people}.    

At work, initiating interactions with new people is often intimidating because of fears of social evaluation~\cite{beck2010looking, eisenberger2003does, boothby2018liking}---people worry that their attempts to initiate interactions may be rejected or that others may not enjoy the conversation~\cite{schroeder2022hello, sandstrom2021people}. \textit{``Will they find it inappropriate?'' ``Will it interest them?''} When initiating a conversation, individuals often worry about the social evaluation they will face~\cite{beck2010looking, eisenberger2003does, boothby2018liking}. Underlying these fears is an uncertainty about other people's interests and what they are willing to talk about~\cite{svennevig1999getting, nguyen2015known}. Awareness of mutual interests enables conversation~\cite{schroeder2022hello, dunn2022can}, however, mutual interests are not always externally perceptible, especially among new acquaintances. As a result, people are often uncertain whether others will accept their attempts to initiate interactions~\cite{sandstrom2021people}, and moreover, whether they will actually enjoy the conversation~\cite{schroeder2022hello, sandstrom2021people}. People naturally avoid situations that can cause embarrassment or social awkwardness~\cite{goffman2008behavior, goffman1967interaction, boothby2018liking}. When common interests are not externally visible, the fear of social evaluation can deter people from initiating interactions with new people altogether~\cite{boothby2018liking}.

What if individuals knew in advance that the people they were about to interact with were interested in conversing about the same topic? This paper introduces a new system, \sys{}, an online tool that supports initial interactions in the workplace by enabling individuals to discover and interact with others they share common interests with, while lowering risks of social evaluation typically involved. When individuals want to find others interested in a topic, they transfer the locus of social evaluation away from themselves by anonymously creating a \textit{nook}---a conversation room that identifies its topic and norms, but not its creator. Before activating a nook, \sys{} probes individuals across the group about their interest in joining it. Finally, by activating a nook to include only those interested in the topic, individuals in a nook are  made aware of their shared desire to interact on the topic. Through this mechanism, \sys{} mitigates concerns of social rejection and shifts the responsibility of ensuring an enjoyable conversation to the group and away from any one individual. Thus, \sys{} attempts to support initial interactions by lowering fears of social evaluation.

By iteratively bringing unacquainted individuals across the workplace in contact with each other over conversations about shared interests, \sys{} attempts to support the development of social relationships at work. \sys{} is an application built for Slack. Underlying it is a continually-running system that probes individuals about their interest in joining various conversations from among an evolving pool of nooks. Anyone across the workplace can anonymously contribute a nook to the pool at any time. \sys{} convenes conversations around shared interests on an ongoing basis by routinely: (1) sourcing and representing topics of conversation as nooks; and (2) convening individuals interested in a nook through scheduled aggregation of individual interests.

To understand, in-situ, how \sys{} can support relationship building among unacquainted individuals at work, we deployed \sys{} for nine-weeks with a group of $25$ \ed{student workers} starting a summer research program. \ed{Participants were full-time paid workers employed as research assistants and worked alongside other research professionals including faculty, research staff, postdocs, and graduate students within the academic workplace.} By analyzing usage data and through interviews with users, we found that \sys{} catalyzed new casual conversations in the workplace by creating an alternate online sphere that provided users with non-threatening and inclusive interaction opportunities. Participants personalized their use of \sys{} to meet their needs for social interaction, with many using it as a way to initiate offline activities. Further, we found that \sys{} distributed the responsibility and ownership of the conversation across individuals in it. In many cases, this led to individuals in a nook collectively driving the conversation forward, however, it also led to some cases where the conversation failed to take off due to social loafing. Finally, we found that \sys{} provided individuals with new awareness about their coworkers interests and desires for social interaction. This often led to conversations outside of \sys{}, and even offline. Based on our findings, and our experience of developing \sys{}, we discuss opportunities and challenges for the design of workplace socialization systems.


\section{Related Work}
Here, we draw on literature across the fields of HCI, organizational behavior, and social psychology to motivate our work. We begin by briefly surveying the individual and organizational impacts of social relationships at work. Then, we examine barriers that inhibit the initiation of interactions. Finally, we situate our work within broader HCI research on the design of social awareness systems.

\subsection{Social relationships at work}
Social relationships at work play a vital role in individual well-being by shaping experiences at work~\cite{baron1996interpersonal, tschan2004work}. Relationships with coworkers have been shown to influence job satisfaction~\cite{cook1981experience, spector1997job}, motivation~\cite{kanfer1990motivation}, and resilience~\cite{dutton2003power}. These relationships can be a source of social support~\cite{elfering2002supportive, tschan2004work}. By serving as vehicles for information, knowledge, and opportunities~\cite{lawler1998network}, connections across the workplace can also accelerate learning~\cite{wenger2000key, lave1991situated}. Further, it is through interactions with others at work that individuals craft and enact their desired work identities~\cite{dutton2003power}, and create pathways for their professional development~\cite{dutton2003power}.

Looking beyond the individual, social relationships within a workplace also have a significant influence \ed{on} the organization as a whole~\cite{baron1996interpersonal}. Social relationships can improve individuals' organizational commitment~\cite{tschan2004work}, organizational citizenship behaviors~\cite{organ1997organizational}, and even result in lower turnover~\cite{george1990understanding}. Taken together, this work suggests that strong social relationships can transform organizations into communities where individuals bring their full selves to work, where collaborators engage each other fully, where learning and development are catalyzed, and where individuals collectively construct and derive positive meaning from their participation in the community at work.

\subsection{Barriers to initiating interactions}
It is by initiating conversations with strangers that people develop the social relationships that ultimately have lasting consequences for their own well-being and the organization as a whole. However, people are often reluctant to start conversations with unknown others~\cite{boothby2018liking, sandstrom2021people}. Initiating interactions with new people is often intimidating because people fear the ensuing social evaluation~\cite{boothby2018liking}. People often worry that initiating interactions with others can cause embarrassment or awkwardness~\cite{sandstrom2021people, boothby2018liking}. They also worry that others may not like them or may not reciprocate their interests~\cite{epley2014mistakenly, boothby2018liking}. The fear of social evaluation in initiating interactions is composed of fears of \textit{trying} to have a conversation with others as well as fears of \textit{actually having} a conversation with others~\cite{schroeder2022hello}.


Fears that relate to \textit{trying} to have a conversation with others, stem from a fear \ed{of} social rejection~\cite{schroeder2022hello}. People worry that others might be disinterested in them or their interests, or may even flat out reject their attempts to initiate a conversation~\cite{schroeder2022hello, sandstrom2021people, epley2014mistakenly}. They worry that the interest they express in others may not be reciprocated~\cite{boothby2018liking, beck2010looking, eisenberger2003does}. However, even if rejection is not a possibility and one expects to actually have a conversation with the person they approach, fears that relate to \textit{actually having} a conversation can cause intimidation. These fears stem from a worry that one's conversation partners will not enjoy the conversation~\cite{sandstrom2021people}. People worry that the conversation might be awkward, unpleasant, or that they may not meet their conversation partners' expectations~\cite{sandstrom2021people, schroeder2022hello}. While work has shown that fears of \textit{trying} to have a conversation are often more intense than fears of \textit{actually having} a conversation, both these fears together create a psychological barrier to initiating interactions.

One of the factors that contributes to these fears is that people lack an understanding of topic preferences of people they are unacquainted with~\cite{nguyen2015known, svennevig1999getting}. Topics of mutual interest are often imperceptible. As a result, people remain uncertain about whether their attempts to initiate interactions will be accepted and whether the conversation will be enjoyable for others~\cite{svennevig1999getting}. Mutual interests can form extended common ground between individuals, driving the conversation forward while also promoting feelings of mutual affinity~\cite{svennevig1999getting, nguyen2015known, clark2006context}. However, when an awareness of mutual interests is absent, people naturally attempt to avoid potential causes of embarrassment or awkwardness (what Goffman calls ``face-threatening acts''~\cite{goffman2008behavior}) and so, they remain reluctant to initiate interactions~\cite{schroeder2022hello}. This can lead to pluralistic ignorance: even when there is a mutual desire to interact,  people may end up not initiating interactions~\cite{epley2014mistakenly, schroeder2022hello}. Motivated by this work, \sys{} contributes the design of online tools and mechanisms that surface common interests and lower the psychological barriers to initial interactions. 


\subsection{Interaction opportunities created by social awareness systems}

Early social awareness systems demonstrated how information about other people's activities can promote social interactions at work. These systems provided users with social signals to keep them informed of what was going on ``around them''~\cite{fitzpatrick1999augmenting}. Dourish and Bellotti describe \textit{awareness} as an ``understanding of activities of others which provides a context for your own activity''~\cite{dourish1992awareness}. Recognizing the role of social awareness in supporting dynamic collaboration at work~\cite{kraut1988patterns, bly1993media, fitzpatrick1999augmenting}, designers of early social awareness systems attempted to enhance community awareness as a means to support collaboration and coordination at work~\cite{erickson1999socially, dourish1992portholes, bly1993media, zhao2000s, yee2005studiobridge, greenberg2001notification}. Towards the goal of supporting coordination in work groups, these systems provided signals such as who is in one's proximity, shared media elements~\cite{greenberg2001notification, cadiz2002designing}, and even what activities are occurring at work~\cite{zhao2000s, kirkham2013break}. Even though most of these systems were not explicitly designed to support social interactions at work, in the contexts in which they were deployed, they were often used for social interactions incidentally~\cite{bly1993media, greenberg2001notification}. 

These systems created opportunities for interactions through awareness about the location and activity of others. However, these systems contributed few signals about shared interests~\cite{zhao2000s}, which can often be crucial for interactions between new acquaintances. Subsequent work has explored how workplace systems can create interaction opportunities between unacquainted individuals by surfacing mutual interests~\cite{nguyen2015known, mccarthy2002using}. For instance, GROUPCAST highlights topics of mutual interest between people who are in close proximity to a shared display~\cite{mccarthy2002using} while other work has explored how such mutual interests can be displayed unobtrusively through Google Glasses~\cite{nguyen2015known}. Both these systems rely on static user profiles to generate topics of mutual interest. However, such profiles can be challenging to create due to concerns of privacy~\cite{mccarthy2002using}, and they can also be unfaithful to the rich, diverse, and evolving interests of the users~\cite{mccarthy2002using}. Relative to this prior work, we do not view interests as static or representable by finite set of fields: \sys{} allows individuals to create conversations about their evolving and contextually-informed interests.

\section{Nooks}
\sys{}, is an online tool that supports initial interactions in the workplace by enabling individuals to discover and interact with others they share common interests with, while lowering risks of social evaluation typically involved. By iteratively bringing individuals across the group in contact with each other over conversations of mutual interest, \sys{} supports relationship building in the workplace. \sys{} is instantiated as a Slack application. We start by describing the underlying building block of \sys{}---a nook---and how it incorporates strategies to lower the risk of social evaluation. Next, we motivate \sys{}' design with an example. We then describe the design and implementation of \sys{} and conclude by describing features that are designed to support adoption and customization.   

\subsection{Nook: A building block to lower the risk of social evaluation}
People are more comfortable initiating interactions when they are aware that the people they are about to interact with, are also interested in the conversation~\cite{schroeder2022hello}. Signals of mutual interests enable conversation~\cite{schroeder2022hello, dunn2022can} however such signals are not always externally visible. As a result, people remain uncertain about whether their attempts to initiate interactions will be accepted (discomforts that relate to \textit{trying} to have a conversation~\cite{schroeder2022hello}) and whether the conversation will be enjoyable for others~\cite{svennevig1999getting}(discomforts that relate to \textit{actually having} a conversation~\cite{schroeder2022hello}). These sources of discomfort stem from an uncertainty about other people's interest in a conversation. \sys{} lowers the risk of social evaluation inherent to initiating interactions, by making mutual interests explicit through its underlying building block--- a nook. 

Whereas individuals are typically subject to social evaluation when they initiate a conversation, a nook shifts the locus of social evaluation away from an individual. A \textit{nook} is a conversation room that identifies its defined focus and norms, but not its creator. Anyone within the workplace can anonymously create a nook about any topic they are interested in. Before being activated, every nook goes through an incubation period, a process during which individuals across the group are probed about their interest in joining it; participants may opt-in or opt-out of conversations as they see fit. Finally, a nook is activated as a short-lived Slack channel adding all those who have expressed interest (including the creator by default), as participants, all at once. 

Because a nook is created anonymously, it transfers the responsibility of identifying interested others, to \sys{}. This has the potential to eliminate discomforts relating to social rejection~\cite{schroeder2022hello, boothby2018liking, sandstrom2021people} because individuals are no longer \textit{trying} to initiate a conversation themselves. By not identifying its creator, a nook places everyone within it on an equal footing. Thus, ensuring an enjoyable conversation~\cite{sandstrom2021people} is no longer the sole responsibility of its creator. This can eliminate discomforts related to \textit{actually having}~\cite{schroeder2022hello} a conversation. Finally, people are only added to a nook if they signal an interest in it and so, everyone within a nook has explicitly signaled their interest in having the conversation---a signal that can eliminate feelings of discomfort in interaction. 

\subsection{Example scenario: initiating affinity groups at work}

Jos\'e has recently joined a new design agency. As someone that enjoys reading mystery novels, Jos\'e is interested in finding others at work with whom he can exchange recommendations and discuss books. He is connected to others through Slack and also has opportunities to approach them in person but he is unsure about who might be interested, and whether people would find his invitation appropriate. So, he decides to use \sys{} to discover others who are interested in mystery novels. To do this, Jos\'e simply creates an anonymous nook on mystery novels (\textit{sourcing and representing interests as nooks}), and in the description, clarifies his intentions to use the space to exchange recommendations and discuss books. Through the incubation period, \sys{}' execution engine accumulates others' interest in joining the conversation (\textit{convening individuals interested in a nook through scheduled aggregation of individual interests}). Finally, at the end of the incubation period, Jos\'e is added to a 24 hour Slack channel with others who have also expressed interest in the nook. Everyone in the conversation knows that they share an interest in mystery novels and so, they feel comfortable exchanging recommendations. They hop in and out of the conversations as they find time, while the nook is active. After their interaction, Jos\'e invites others in the nook to a group chat as a space for recurring interactions.


\subsection{System design}
\sys{} is a Slack application that manages a dynamic pool of nooks, to which anyone can anonymously contribute a nook that they would like hosted. A continually-running engine incubates nooks by routinely making incubating nooks visible to individuals across the group and probes them about their interest in each. Here, we describe the features of the system, focusing on: (1) how users can make ongoing contributions to the pool of nooks through the \sys{} homepage (Figure \ref{fig:interface}); and (2) how the continually-running engine routinely probes individuals' interest and activates nooks, with the flow shown in Figure \ref{fig:flow}.
\subsubsection{Sourcing and representing topics of conversation as nooks}
\begin{figure*}[t]
    \centering
    \includegraphics[width=\textwidth]{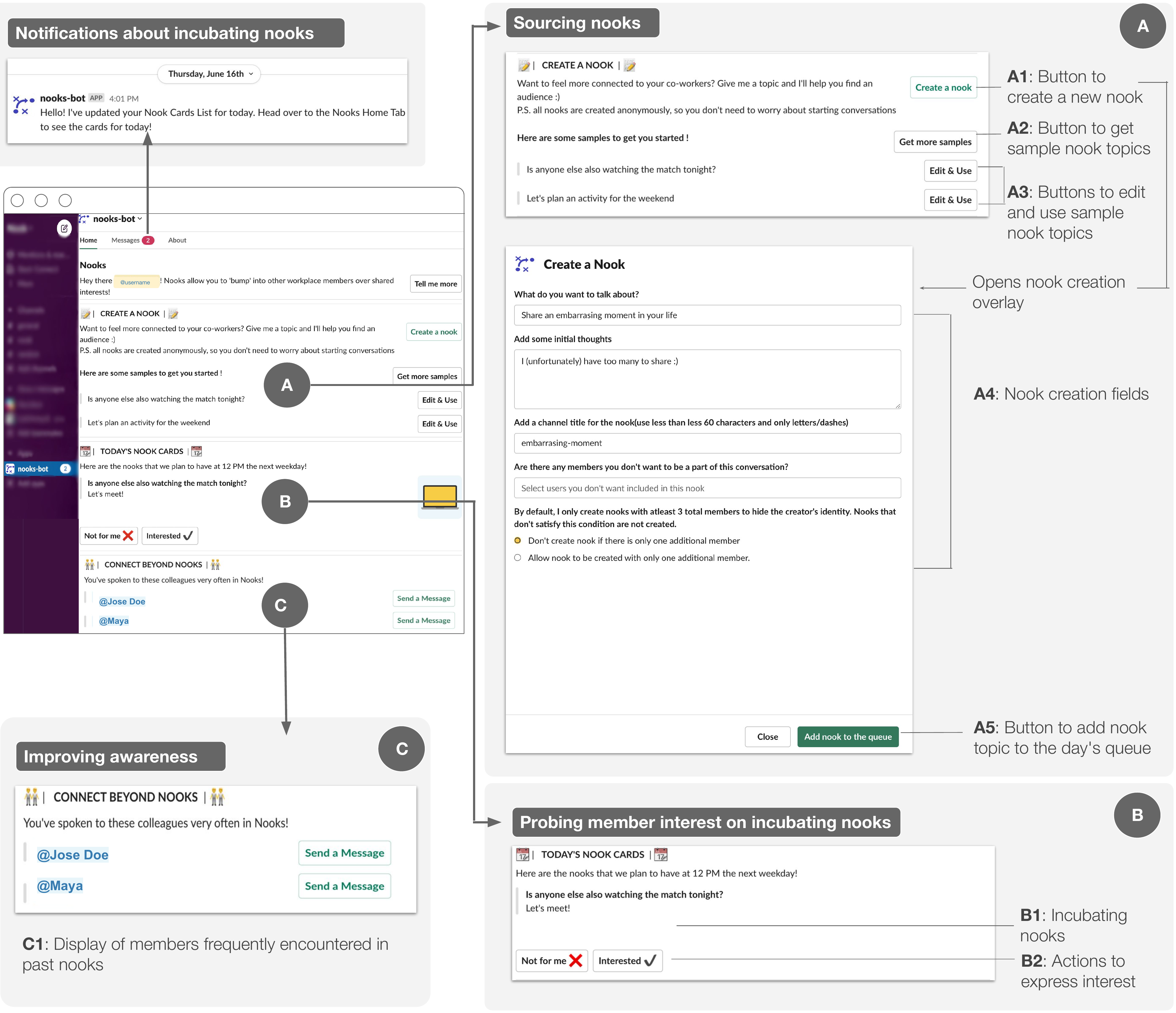}
    \caption{A tour of the application homepage. The `create a nook' panel (A) allows users to anonymously create a nook (A1), as well as edit and use sample nooks (A2,A3). Attempting to create a nook, opens up the nook creation overlay (A4) that asks users to provide a title for their nook as well as their description of what they want to talk about. The homepage also shows users nooks that are currently being incubated (B), displaying the topic and description of the nook (B1) and providing them with the choice to opt-in/opt-out (B2). To alert users of available nooks, the application also sends notifications to users in the form of a Slack direct message. Finally, the homepage also shows users a list of other users they encounter most often in nooks (C1).}
    \Description{“A tour of the application homepage. Different parts of the Nooks Bot homepage are shown here. The image is broken up into four main sections to focus on these aspects separately: Sourcing Nooks, Probing member interest on incubating nooks, Improving awareness, and notifications. The first section is demarcated as "Create a Nook" in the Nooks bot homepage. The section includes a button to create a nook(A1), a button to get more samples (A2) and two sample nook topics with an Edit and Use button next to them(A3). The two samples shown in the image are: "Is anyone else also watching the match tonight", and "Let's plan an activity for the weekend". The "Create a Nook" button is shown to open an overlay form with the following questions(A4):  "What do you want to talk about?": this field requires a text input and the pre-filled input in "Share an embarrassing moment in your life"."Add some initial thoughts": this field also requires a text input and the pre-filled input in the image is "I (unfortunately) have too many to share :)"."Add a channel title for the nook"(use less than 60 characters and only letters/dashes): requires a text input and the pre-filled input in the image is "embarrassing-moment"."Are there any members you don't want to be a part of this conversation": this field requires users to select from options and the helper text given is "Select users you don't want in this nook". The bottom of this page has a "Add nook to the queue" button(A5). The second section in the home page is "Today's Nook Cards" followed by a nook title and description(B1). The specific sample given is "Is there anyone else also watching the match tonight?" with a description of "Let's meet". This is followed by two buttons titled "Not for me" and "Interested"(B2). The third section improving awareness: the section is called "Connect beyond Nooks" on the homepage and includes a list of sample users(in the image: @John Doe, and @Maya) with corresponding "Send a message" buttons(C1). The fourth section in the image shows the notification send out by the application. It appears as a Slack DM and reads: “Hello! I've updated your Nook Cards List for today. Head over to the Nooks Home Tab to see the cards for today!”}
    \label{fig:interface}
\end{figure*}
For \sys{} to bring people in contact with each other over common interests, it needs a representation of people's interests that it can then use to connect individuals. What people want to talk about with others is often context specific and evolving, varying with the nature of interpersonal relationships as well as surrounding context. To capture individuals' evolving interests in a way that can be used to attract like-minded others, \sys{} lets users anonymously create nooks whenever they want, by providing a topic for the nook and a description of what they want to talk about within it.
To create a new nook, users can proceed to the ``Create a Nook'' panel in the Nooks home page (Figure \ref{fig:interface}) within Slack. Clicking on ``Create a Nook'' (A1) opens an overlay form (A4) asking the user to describe the nook they want to create. Specifically, they are asked to provide a `title' and their `initial thoughts'. Here they can describe not just \textit{what} they want to talk about but also what they want the norms of the conversation to be. By default, nooks are open for anyone in the group to join, however, the creator can optionally select specific users they don't want included--- a control requested by many users in our pilot studies. Finally, they can create the nook by clicking the ``Add Nook to the Queue'' button at the bottom of the overlay(A5). This panel also shows users a variety of pre-populated sample nooks to inspire users. They can navigate through these samples by clicking on ``Get more samples'' (A2). Clicking the ``Edit and Use'' button (A3) opens the overlay form (A4).

\subsubsection{Convening individuals interested in a nook through scheduled aggregation of individual interests}

Using a user-contributed nook to then assemble a group of people interested in it requires a mechanism that probes people across the workspace about their interest in joining it. Within \sys{}, each user-contributed nook is ``incubated''---other users across the group are shown the topic and description of the nook and asked for their interest in joining a nook. Next, we describe how nooks are incubated and activated.

\begin{figure*}[t]
    \centering
    \includegraphics[width=\textwidth]{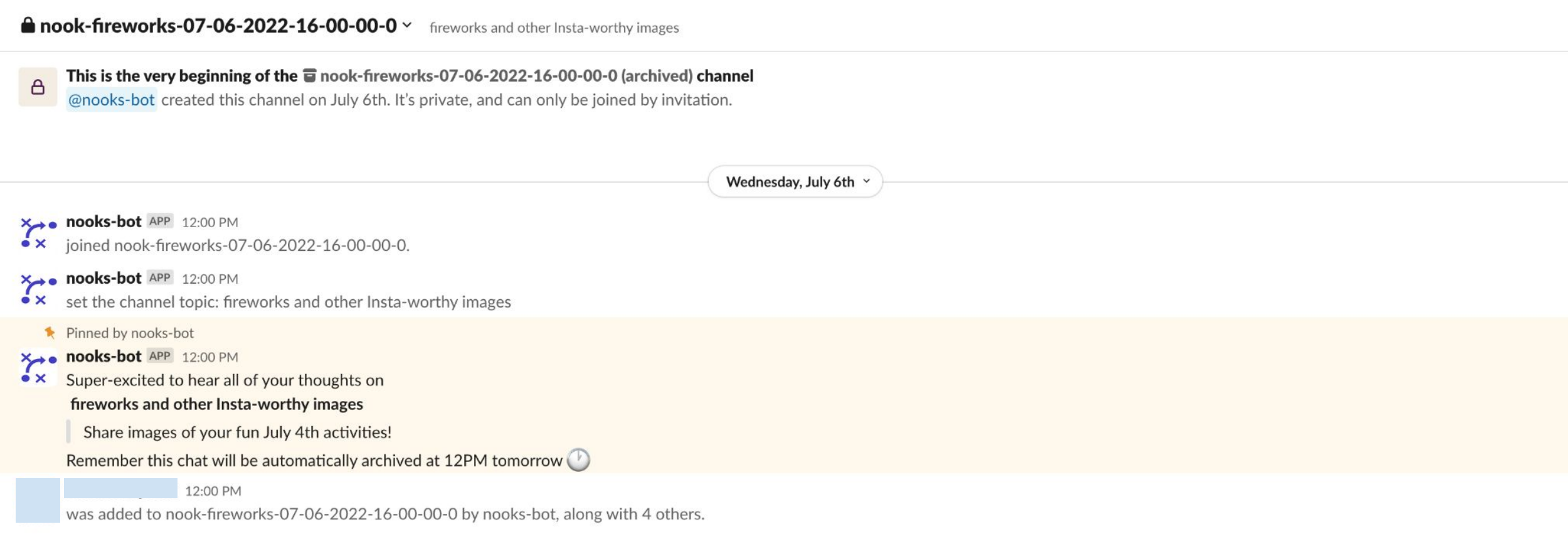}
    \caption{At the end of the incubation period, each nook is activated as a private Slack channel to which the \sys{} bot automatically adds only those users who had expressed interest, including the creator. The bot greets the channel by posting the topic and the initial thoughts added by the creator but does not identify the creator.}
    \Description{The image shows a Slack channel. The nooks bot is a member of the channel and has set the topic of the channel to be “fireworks and other Insta-worthy images”, followed by a message from the bot that reads: “Super-excited to hear all of your thoughts on fireworks and other Insta-worthy images. Share images of your fun July 4th activities! Remember this chat will be automatically archived at 12PM tomorrow”. The figure shows that five members have been added to the channel.}
    \label{fig:nook}
\end{figure*}

\textit{Incubation and activation:} When a nook is being incubated, it is displayed on the \sys{} homepage (Figure \ref{fig:interface}, panel B) for all members across the workspace, excluding those who the creator has explicitly requested to exclude. Users can click `interested' or `not for me'(B2) on each nook they are shown (B1). If multiple nooks are being incubated at the same time, they are shown sequentially. \sys{} records each user's choice. \ed{When asking users about their interest in joining an incubating nook, we only display the `title' and `initial thoughts' but do not display any social signals such as how many others have expressed interest. Social signals can cause people to conform to the choices of others~\cite{milgram1969note, cheng2014catalyst} instead of choosing nooks based on what they personally find interesting. This can undermine \sys{}' goal to match people to conversations based on their interest in the conversation and so, we exclude social signals to ensure that users' decisions about which nooks to join are primarily driven by their own interest in the nooks' topics and not by others' choices.} At the end of the incubation period, each nook is activated as a private Slack channel (not discoverable to non-members) to which the \sys{} bot automatically adds only those users who have a priori expressed interest. The bot greets the channel by posting the topic and the initial thoughts added by the creator but does not identify the creator. Figure \ref{fig:nook} shows an example of a nook once activated. \ed{Launching of nooks is in no way conditioned on their popularity. Online spaces that optimize for activity attempt to minimize sparsely populated spaces in favor of more populous spaces~\cite{10.7551/mitpress/8472.003.0007}. This perspective would recommend launching only the most popular nooks. However, \sys{} focuses not on maximizing activity, but on supporting relationship building which can occur just as well, if not more effectively, in smaller conversations: smaller conversations are more effective at fostering openness and trust~\cite{lowry2006impact, soboroff2012group}}.

Slack channels that host nooks are always activated at 12pm and open for 24 hours, after which they are automatically archived. \ed{Deciding the lifetime of conversation spaces represents a trade-off between supporting engaging conversation and minimizing work disruption. Long-lived spaces, and especially persistent spaces, can lead to diffused participation and can impede conversation. Best practices on the design of online spaces recommend establishing `expected active times', during which participants can expect each other to be active in the space~\cite{10.7551/mitpress/8472.003.0007} and consequently, promotes activity in a self-fulfilling way. Since users may not always be active on Slack, and diffused participation can impede conversations, limiting the lifetime of nooks can help establish expected times of activity and promote engaging conversations. However, conversation spaces that are too short-lived (such as synchronous channels) can heighten attentional demands on individuals and disrupt work performance by inviting individuals to participate when they would prefer to focus on other tasks. Setting the lifetime of a nook to be 24 hours represents a middle ground, balancing the opposing objectives of enabling engaging conversation and minimizing work disruption---it hopes to achieve the benefits of establishing `expected times of activity' while minimizing work disruptions as individuals can participate whenever they find time during the 24 hour period.} Participants are notified when the conversation is archived and if participants in a conversation want to convert the nook into a more persistent space, they have the option to unarchive the channel. We chose to activate nooks at 12pm because that was the time by which most people in our pilots had started their workday, however, this time can be customized to a workplace depending on the distribution of times when individuals start their workday.

\begin{figure*}[t]
    \centering
    \includegraphics[width=\textwidth]{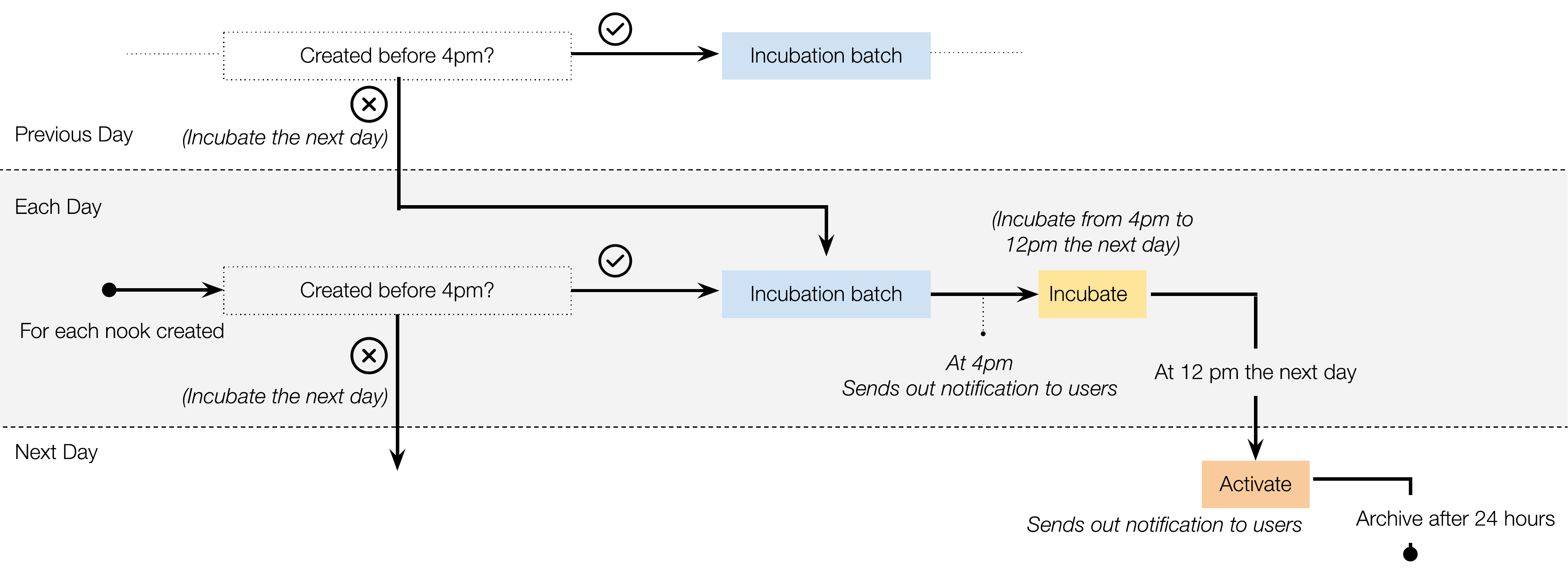}
    \caption{The flow of a nook through the execution engine: All nooks created before 4pm on any given day are added to that day's incubation batch and are incubated together from 4pm that day to 12pm the following day. Users are notified that a new batch of incubating nooks is available via a Slack message triggered by the application. Nooks being incubated together are displayed sequentially to users on the homepage. At 12pm the following day, incubating nooks are activated as Slack channels, including all members who have expressed interest in it by then. Nooks created after 4pm are added to the following day's incubation batch.}
    \Description{A flow chart that shows the time-based decisions made about whether to incubate a nook on the given day or the subsequent day. Following this, the diagram shows when a nook is incubated, when it is activated, and when it is archived. Additional details are provided in the caption.}
    \label{fig:flow}
\end{figure*}
\textit{Incubation and notification routine:}
\sys{} houses communities at the scale of a workplace and so, interaction opportunities---incubating nooks---within it can be sparser than interaction opportunities, such as discussion threads, in large online communities. Prior work suggests that when interaction opportunities are sparser, users lower the frequency of their visits~\cite{10.7551/mitpress/8472.003.0007}, which can result in some incubating nooks going unnoticed. To ensure that incubating nooks are viewed by a sufficient number of users, we chose to use a \textit{push model} where users are notified when new interaction opportunities---incubating nooks---are available~\cite{10.7551/mitpress/8472.003.0007}. However, notifying individuals every time a nook is created can lead to excessive notifications. To simultaneously ensure that each \ed{nook} is viewed by a sufficient number of people and that the interruption cost of notifications is low, we devised a temporal routine according to which nooks are incubated in batches. \ed{Batching information interruptions can minimize their negative impact on productivity~\cite{mark2016email}}. Figure \ref{fig:flow} shows this routine. All nooks created before 4pm on any given day are added to that day's incubation batch and are incubated together from 4pm that day to 12pm the following day. Users are notified that a new batch of incubating nooks is available via a Slack message triggered by the application (Figure \ref{fig:interface}). Nooks being incubated together are displayed sequentially to users on the homepage. At 12pm the following day, an incubating nook is activated as a Slack channel and includes all members who have expressed interest in it by then. Nooks created after 4pm are added to the following day's incubation batch. 

\subsection{Implementation}
\sys{} is a Slack application and a companion Slack bot, implemented in Python with a Flask back-end\ed{\footnote{\href{https://flask.palletsprojects.com/en/2.2.x/}{https://flask.palletsprojects.com/en/2.2.x/}}}. We used the Slack Bolt API\ed{\footnote{\href{https://api.slack.com/tools/bolt}{https://api.slack.com/tools/bolt}}} to create the Slack interface and monitor events that are triggered as users interact with the bot. The back-end was served on a Digital Ocean\ed{\footnote{\href{https://www.digitalocean.com/}{https://www.digitalocean.com/}}} instance with a MongoDB database\ed{\footnote{\href{https://www.mongodb.com/atlas/database}{https://www.mongodb.com/atlas/database}}}. To maintain users' privacy, we do not log any conversations. We collect users' demographic data when they create their profile. Additionally, for every nook created, we record the title and details of the nook created, the user ID of the creator, the user IDs of those who express interest and disinterest in the nook, and finally user IDs of all members who are added to the nook. When a user first signs-up on the application, they are walked through the consent form within the application. The project is open source and available at: \href{https://github.com/Sbali11/Nooks}{https://github.com/Sbali11/Nooks}.

\subsection{Adopting, customizing, and promoting use within a workplace}
We envision \sys{} as a tool that workplace administrators and decision-makers---those responsible for making workplace-wide decisions---can adopt for their workplace as a way to support relationship building in the workplace. Like most social computing systems, for \sys{} to be useful as a tool for initial interactions, individuals across the workplace need to sign-up as and actively use the application. Once installed in the Slack workspace of the workplace, \sys{} allows administrators to onboard members from a specific channel in the Slack workspace, to control who participates in \sys{}. Onboarding members from the \textit{\#general} channel, for instance, would onboard everyone in the Slack workspace. Alternately, they can also invite specific members using their Slack usernames. \sys{} sends invited users a Slack message, walking them through the sign-up process. 

Administrators can further customize \sys{} to their workplace by editing the sample nooks (A2 in Figure \ref{fig:interface}) to influence the nature of conversations. For instance, they might insert sample nooks relating to a recurring event, if they want to promote conversations about it. Finally, to promote the use of the application, administrators can create a predefined set of nooks that are inserted into the application during set-up. Allowing users to view and join these predefined nooks, provides users with interaction opportunities even before they have created their own nooks. When populated with predefined nooks, by administrators, \sys{} acts similar to systems that deliver online icebreaking prompts to groups. However, they differ in that prompts are only \textit{suggested}, not \textit{enforced}: users make decisions about what conversations they join. Participating in predefined nooks also supports users in learning \sys{}' underlying mechanism and allows them to experience conversations in a nook which can inform their decision to create their own.

\section{Deployment Study}
We conducted a nine-week deployment study to investigate how people used \sys{} in the field. To maximize ecological validity, we asked participants to use the application however, and as frequently as they desired, and observed patterns of use that emerged naturally. There was no monetary compensation associated with using the app. (This set up is different from previous studies in the workplace, such as those focused on workplace connection and engagement~\cite{williams2018supporting}, which compensate users for using the intervention.) This study was approved by our university’s Institutional Review Board.

Specifically, this longitudinal deployment focused on these research questions:
\begin{itemize}
    \item RQ1---\textit{How do participants perceive the experience of interacting within a nook?}
    \item RQ2---\textit{How do participants use \sys{}?} What kinds of communication does it afford and what patterns emerge?
    \item RQ3---\textit{How does \sys{} influence relationship building in workplaces?}
\end{itemize}

\subsection{Participants and Research Setting}
Given our focus on supporting relationship building in workplaces, we worked with administrators of a summer research program at a private university in the United States for this study. At the start of the program, a total of $25$ students ($19$ female, $4$ male, and $2$ non-binary) in the program joined the deployment on a voluntary basis. All participants were aged 18-24. \ed{As part of the summer research program, students were employed full-time during the summer and paid a salary as research assistants. These student workers were primarily engaged in conducting research alongside existing research professionals at the university including faculty members, postdocs, and graduate students. Participants worked in areas relating to computing including smart classroom sensors, educational games, accessibility, and smartphone privacy tools. None of them were taking classes or engaged in other part-time positions. Student workers participating in our study were essentially new employees in this academic workplace. Further, only 20\% of the students in the program were affiliated with the university prior to the start of the employment and so, 80\% of the students were new to the university as a whole.} Participating students had desk assignments across four different buildings across the university campus but were within walking distance of each other. \ed{This was their primary work area.} Students were fairly mobile, occasionally choosing to work from home. \ed{Student workers were all located in the same city and they interacted with each other both in-person and online.}

The program design included some socialization: administrators for the summer research program had already added all the students to a common Slack workspace to create a space for announcements, while also giving them a space to interact with each other. In addition, as part of the program, students were also invited to attend a seminar session twice a week where invited speakers would present their work to the students. \ed{Beyond this, since students shared work areas, opportunities to initiate interactions would also arise opportunistically when they would bump into each other.} Because students were just starting out in the program, they were also new to each other and this provided a valuable context to study the use of \sys{} and how it might support early stages of relationship building.

\subsection{Procedure}
We first worked with the administrators of the program to add the \sys{} application to the Slack workspace used by the students. Then, we held a short demonstration session for the students during an upcoming weekly seminar. This demonstration session comprised a brief tutorial on how to sign-up on \sys{}, how students could contribute their own nooks, express interest in contributed nooks, and related functions. 

We informed students that \sys{} was part of a research study and we wanted to learn about their experience using it. We also informed students that if they chose to sign-up, they would be enrolled as participants in our study and that doing so would allow the application to monitor their usage levels for nine weeks, but that the application did not record the content of any conversation. At the end of the demonstration, the program administrators sent all students in the workspace an invitation through Slack to sign-up on the application. To encourage naturalistic behavior, there was no monetary incentive for signing up on or using \sys{}. Participants were free to use the application as they wanted and could discontinue their use or deactivate the application at any point. 

To bootstrap usage, we populated the application with $9$ predefined nooks (Figure \ref{fig:default-nooks}) which were basic icebreaking prompts. These nooks were incubated on different days during the first three weeks only. In later weeks, only participant-created nooks were incubated. To capture their usage of \sys{}, we recorded the nooks created by participants, as well as the nooks in which they expressed interest or non-interest.

After nine weeks, we invited all participants to additionally participate in a semi-structured interview about their experience using \sys{}. Nine of the $25$ participants accepted the offer to be interviewed and were compensated \$15. Interviews were semi-structured and lasted between 20 and 35 minutes. Interviews were conducted in English by one of the authors. Interviews focused on the participants' overall thoughts and perceptions about the application, their experience of participating in nooks, and how they used the application. To help participants recall their experiences, participants were encouraged to open the application and revisit their conversations within \sys{} while responding to questions. All interviews were recorded and transcribed.

\subsection{Analysis}
Our findings are based on a triangulation of (1) usage data of the participants including the nooks participated in, nooks created, and their expressed preferences among nooks; and (2) transcripts of the recorded interviews. One of the authors performed an initial line-by-line open-coding of the transcripts, iterating over the transcripts as necessary. Codes generated in this phase were in part inductive, driven by the data, and in part guided by our original research questions-- we remained open to capturing observations that emerged through the data while also looking out for observations that related to our main guiding questions. Finally, all authors collectively discussed the analysis and iteratively generated, refined, and solidified themes. Themes were generated at a semantic level, reflecting what participants explicitly said~\cite{braun2006using}.

\section{Findings}
\begin{figure*}[t]
    \centering
    \includegraphics[width=\textwidth]{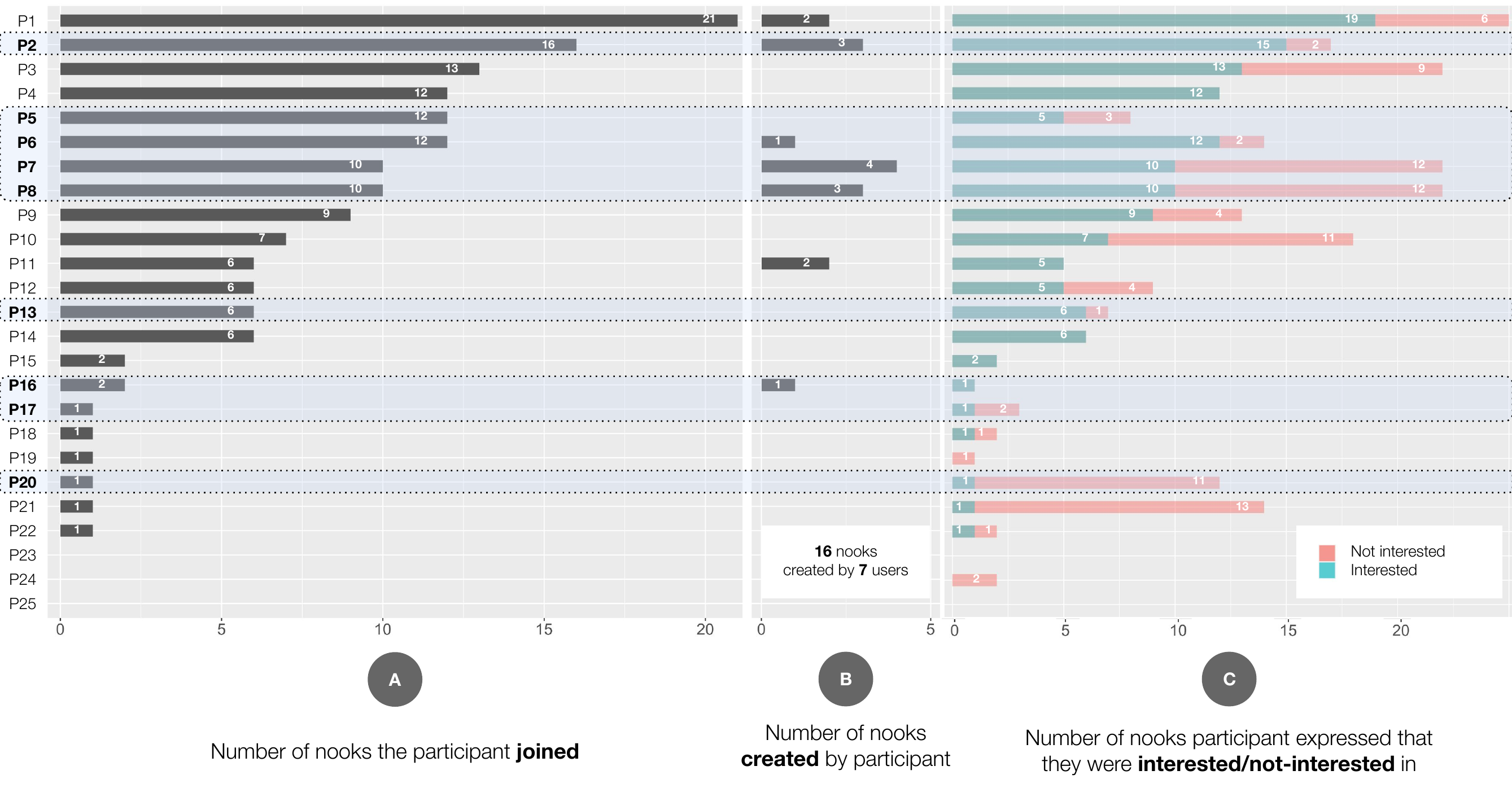}
    \caption{Summary of participants' engagement with \sys{}. Participants who agreed to be interviewed are highlighted in blue. A) The $25$ participants varied in how many nooks they joined. The median number of nooks joined was $6$. Additionally, $22$ participants participated in at least one nook. B) $7$ participants contributed a total of 16 nooks C) Participants also varied in how picky they were about selecting nooks to join. \ed{In some cases, the number of nooks participants interacted in were more than the number of nooks they expressed interest in (eg. P12, P19) because they were manually added to those additional nooks by interactants in those nooks.} Differences in the number of nooks that participants responded to also reveals varying levels of use and monitoring of the application. \ed{Responses of non-interest communicate that users viewed the incubating nook but did not want to join it, indicating a low propensity for those specific nook topics. On the other hand, an absence of responses indicates that users did not see the nooks homepage or did not view incubating nooks on certain days indicating a lowered desire to explore incubating nooks on some days.}}
    \Description{The image shows three separate histograms. The first histogram shows bars corresponding to the number of nooks each participant joined. This histogram is labeled “A”. The second histogram shows bars corresponding to the number of nooks each participant joined. This histogram is labeled “B”. The third histogram shows stacked bars to represent the number of nooks each participant expressed that they were interested/not interested in. This histogram is labeled “C”. Participants are arranged in descending number of nooks they joined in and labelled from P1-P25. Participants P2, P5, P6, P7, P8, P13, P16, P17 and P20 are highlighted.}
    \label{fig:participant-stats-1}
\end{figure*}
\begin{figure*}[t]
    \centering
    \includegraphics[width=\textwidth]{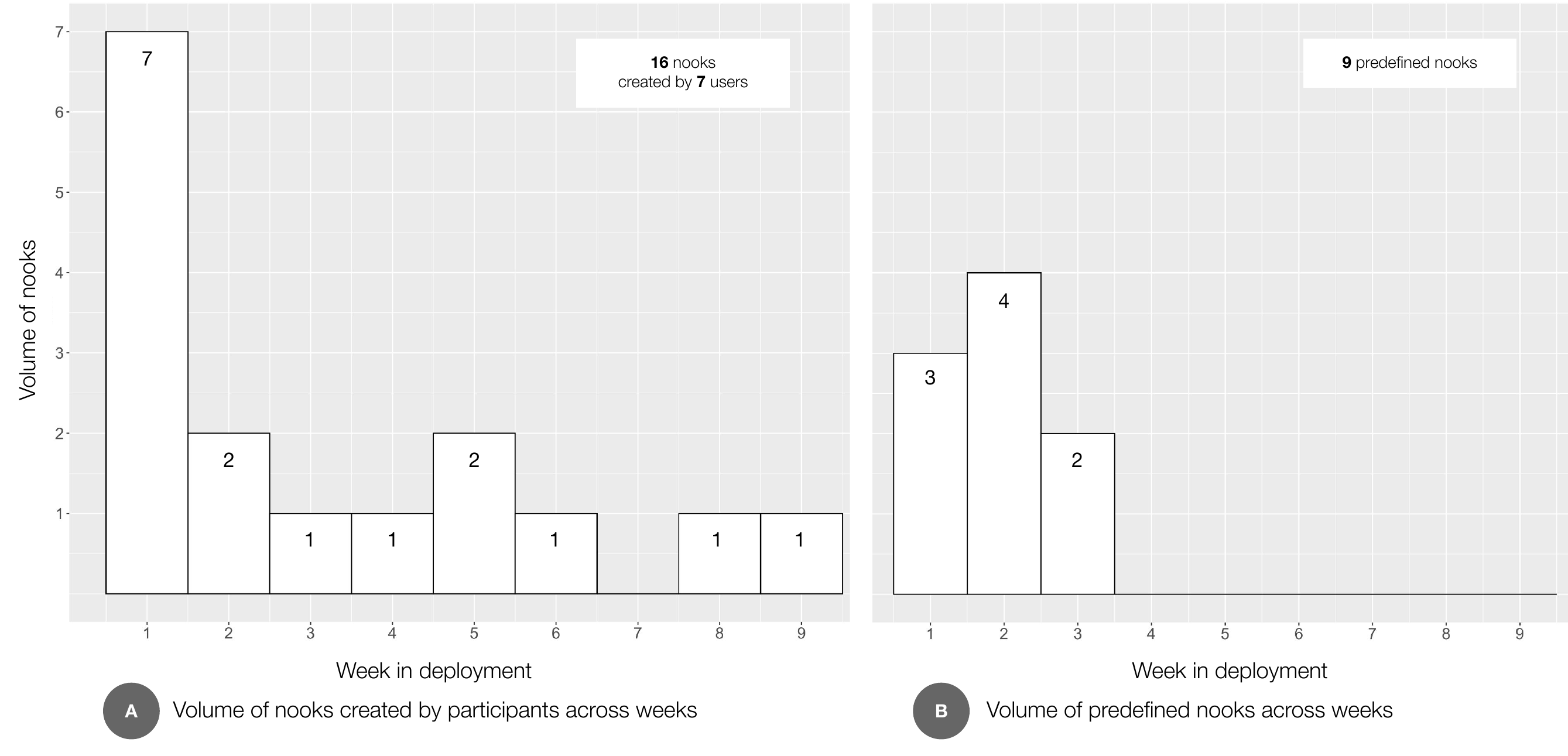}
    \caption{Number of nooks activated within the application across the weeks of deployment varied. A) User created nooks, though present most weeks, were concentrated towards the start. B) Nine predefined nooks (that were inserted to bootstrap use) were only activated during the first three weeks.}
    \Description{The image shows two histograms. The first histogram shows the volume of nooks created by participants across the weeks of deployment. Week 1 witnessed the largest number of pre-defined nooks (7) and in subsequent weeks, the number of nooks ranged from 1-2. The second histogram shows the volume of predefined nooks across the weeks in deployment. Pre-defined nooks were only created in the first 3 weeks.}
    \label{fig:participant-stats-2}
\end{figure*}
\begin{figure*}[t]
    \centering
    \includegraphics[width=\textwidth]{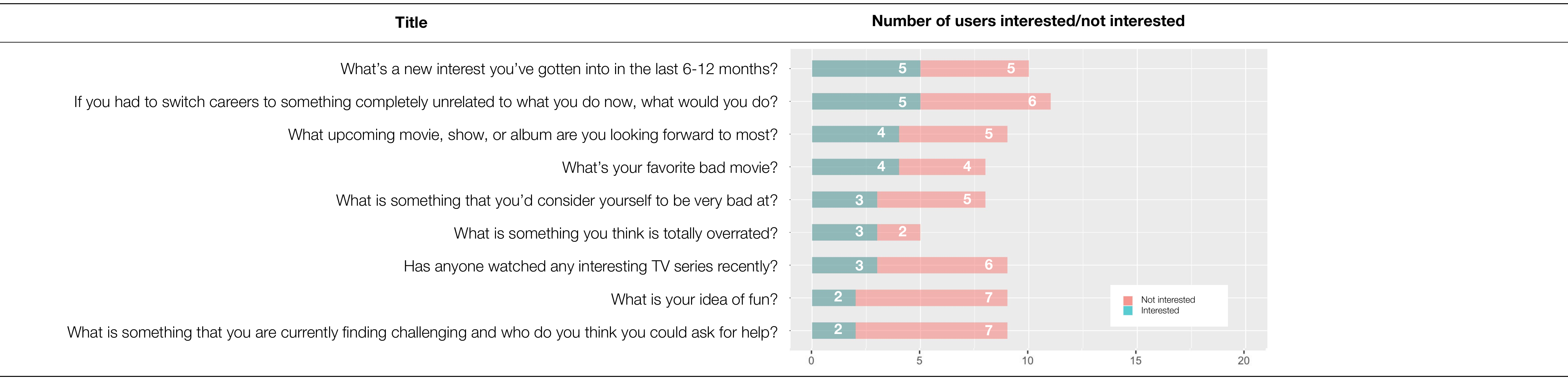}
    \caption{Predefined nooks were basic icebreaker prompts intended to catalyze initial use. These nooks did not have a description attached to them---only a topic. $9$ predefined nooks were activated during the study period and had between $2$ and $5$ participants each, with a median of $3$ participants. \ed{Here, the total number of responses against each nook indicates the number of participants that viewed and responded to the nook when it was incubating.}}
    \Description{The figure shows a histogram with bars corresponding to each predefined nook. The bars are arranged top to bottom in descending order of the number of members they had.  From top to bottom, the nooks read: What’s a new interest you’ve gotten into in the last 6-12 months? If you had to switch careers to something completely unrelated to what you do now, what would you do? What upcoming movie, show, or album are you looking forward to most? What’s your favorite bad movie? What is something that you’d consider yourself to be very bad at?What is something you think is totally overrated? Has anyone watched any interesting TV series recently? What is your idea of fun? What is something that you are currently finding challenging and who do you think you could ask for help?}
    \label{fig:default-nooks}
\end{figure*}

Over the $9$ week study period, participants used \sys{} to have $25$ conversations. $16$ of these conversations were on topics contributed by $7$ participants (Figure \ref{fig:participant-stats-1}). Each of these conversations had between $4$ and $15$ participants, with a median of $9$ participants (Figure \ref{fig:user-created-nooks}). The remaining $9$ nooks were predefined nooks and had between $2$ and $5$ participants each, with a median of $3$ participants (Figure \ref{fig:default-nooks}). Levels of use varied across the participants and across the $9$ weeks of the deployment. $22$ participants participated in at least one nook (Figure \ref{fig:participant-stats-1}) \ed{and the median number of nooks that participants joined was $6$}. Activity was concentrated towards the initial weeks of the deployment and decreased subsequently (Figure \ref{fig:participant-stats-2}). \ed{Although every incubating nook was displayed on every paticipant’s \sys{} homepage, the frequency with which individuals viewed the \sys{} homepage, and responded to incubating nooks (regardless of whether the response expressed interest or non-interest), varied across individuals and across time. As a result, the number of incubating nooks that each individual responded to varies across individuals (Figure \ref{fig:participant-stats-1}(C)), and the number of users that viewed and responded to each incubating nook also varies across nooks (Figure \ref{fig:user-created-nooks} and Figure \ref{fig:default-nooks}). None of the $25$ participants in the study formally withdrew participation during the course of the study, however, like previous deployments of social computing systems, we did observe non-use~\cite{laumer2017challenge, bernstein2010enhancing}, especially $3$ participants (P23, P24, and P25) who did not participate in even a single nook (Figure \ref{fig:participant-stats-1}(A)). 

We invited all $25$ participants to additionally participate in interviews and $9$ participants agreed to be interviewed.} Those who participated in the interviews varied widely in their usage of \sys{} (highlighted in blue in Figure \ref{fig:participant-stats-1}) and included $8$ female and $1$ male participant. \ed{Participants in our interviews represent a reasonably stratified sample---$5$ interviewees participated in more nooks than the median user ($> 6$ nooks) while $4$ participated in fewer than or as many nooks as the median user ($\leq 6$ nooks).

Next, we present the major themes that we identified in our analysis, each presented as a separate subsection.  We first discuss how \sys{} provided participants with a sociable online sphere that lowered psychological barriers to initiating interactions, and fostered feelings of inclusivity in conversations (Section \ref{sec:theme-1}). Then, we discuss how \sys{} distributed the responsibility and ownership of a conversation across all individuals in a conversation and how this affected conversations (Section \ref{sec:theme-2}). Together, these findings speak to participants' experiences interacting within nooks (RQ1). We next discuss how participants' use of \sys{} was shaped by their socialization needs, with many using it to initiate offline activities (Section \ref{sec:theme-3}). These findings describe emergent patterns of use (RQ2). Finally, we discuss how interaction opportunities and improved awareness that resulted from \sys{}, supported relationship building (Section \ref{sec:theme-4}), illuminating the ways in which \sys{} might influence relationship building (RQ3).}

\subsection{\sys{} provided a sociable online sphere that catalyzed new casual conversations}
\label{sec:theme-1}

\subsubsection{\sys{} provided non-threatening ways to initiate and engage in casual conversations}
Participants mentioned how typical channels to engage in interactions online---such as the large broadcast channels in Slack---were not conducive to initiating interactions because they require individuals to address the group as a whole, which could be intimidating (P5, P13, P16, P17). \sys{} lowered this barrier, for participants, by providing them with a non-threatening way to initiate and engage in such initial interactions. For instance, P13 mentioned:
\begin{quote}
I think it makes it a lot easier to do online conversations, because I guess we don't really have anything to start a conversation with online, especially in a Slack channel where you have, everyone in there. It's kind of awkward to just start a conversation with a 60 person slack so I mean it definitely made it easier there. (P13)
\end{quote}

Similarly, P16 mentioned how \textit{"talking to everybody, as a collective can be a bit overwhelming''} while P17 called this a \textit{``psychological burden''}. Because \sys{} allowed people to find others through anonymously-created topics, participants suggested that \sys{} provided a non threatening approach to initiating interactions:
\begin{quote}
I think it's less intimidating and it's also anonymous so I think people who might not be super extroverted or [those uncomfortable] just throwing an idea out there [to the whole group], would feel most comfortable [using Nooks]. Because you can actually gauge whether people are interested or not...and most of the time, at least a couple of people will be interested in your topic. I think it's less intimidating and can really help when you're new to a workspace. (P5)
\end{quote}

By establishing that a topic was acceptable to talk about within a nook, and that everyone within the nook had opted into this norm, \sys{} eliminated P17's hesitation about engaging in casual conversations with others. Other participants described how \sys{} contributed an \textit{``informal environment''}(P2, P5) and provided a venue for conversations that are less ``important'' than you might post in the larger channels (P2), suggesting that \sys{} provided an alternate sociable sphere for casual conversations. Additionally, because conversations within \sys{} usually involved groups of people, it increased participants' levels of comfort going into conversations (P7,P8). P8 noted, \textit{"you know going into the group, you have more than one other person you're going to be interacting with and so it's not going to be this awkward one on one thing''}.

\subsubsection{\sys{} fostered feelings of inclusivity within conversations}
By bringing people in contact with others over shared interests and making the shared interest the focus of conversation, \sys{} promoted inclusivity by making pre-existing ties less pronounced within conversations. P17, who joined the program late, mentioned how she had fewer opportunities to initiate interactions with others because they had already formed tight-knit groups:\textit{"I still remember in our first seminar I just came and stayed in the last row. I didn't know who to talk to, and even people who were sitting next to me seemed like they didn't really want to talk to me and were just talking to people they already knew.''} P17 mentioned how \sys{} gave her a chance to get to know others better, reflecting on how otherwise she might have just known what others in the group \textit{"look like and probably their name.''} P16 also mentioned how \sys{} contributed feelings of inclusivity by naturally including everyone interested in a conversation:  
\begin{quote}
Creating a channel and adding a bunch of people intentionally is bound to create some conflict, if you forget a person or something so I thought [Nooks] was a low effort way to allow other people to opt in.
\end{quote}

In typical chat conversations, everyone is excluded by default: the creator is required to invite people they want to the conversation. In contrast, \sys{} includes everyone by default, while providing creators with controls to exclude members. In doing so, \sys{} naturally prioritizes inclusivity. While this is the very mechanism that makes them useful for initiating interactions, they can often be ill-suited for interactions where the initiator wants to include only specific individuals in the conversation. For instance, P2 recognized how nooks takes away some control over who participates in a conversation: \textit{"[nooks] can only be sent to the entire group and you can't decide who goes in''}.

\subsubsection{Visible activity in \sys{} encouraged participation}
Existing activity within \sys{} stimulated participants to engage within it, creating a kind of honeypot effect~\cite{wouters2016uncovering}. P5 mentioned how others' use of \sys{} prompted her to participate and use it as a way to find new connections : \textit{"I think it definitely encouraged me to see others using it and trying to find each other with Nooks''}. Participants also mentioned how predefined nooks reminded them that they could contribute their own conversation topics and stimulated further use (P5,P8).


\subsection{A nook was everyone's responsibility -- but occasionally, no one's responsibility}
\label{sec:theme-2}
Through our interviews, we found that \sys{} distributed the responsibility and ownership of the conversation, across the group, unburdening creators' of the pressure of social evaluation (P2, P6, P7, P8, P16). Thus, the onus of driving the conversation forward was on everyone in a nook and not just the creator (P2, P6, P7, P8, P16). As a result of this, in some cases, conversations gained momentum even without the creator driving the conversation. P16 mentioned how by the time she joined the conversation in a nook she had created, the conversation had already taken off: \textit{"I think I was busy or doing something in the hour that it was created and by the time I checked my messages someone had already messaged the chat and there was a conversation going on. So it indicated to me that people are actually interested in this. I'm glad I created it.''} 

However, because the creator was no longer solely responsible for the conversation, it occasionally led to failures through social loafing, where the conversation failed to take off as no participant drove the conversation. P8 mentioned how even though joining a nook communicated intention, there were still situations where people remained inactive within a nook. P6 and P7 suggested that nooks were more likely to succeed when the creator made the effort to steer the conversation. P7 mentioned:
\begin{quote}
I feel it would be nice for the person who created the nook to start the conversation or to have a description of what they want to say first and have that be in the nook when you join, so the conversation has already started and you don't have to wait for someone.
\end{quote}

\subsection{Participants' needs for socialization shaped how they used \sys{}}
\label{sec:theme-3}
The extent to which and the ways in which participants used \sys{} were reflective of their desire for social interaction with the group.

\begin{figure*}[t]
    \centering
    \includegraphics[width=\textwidth]{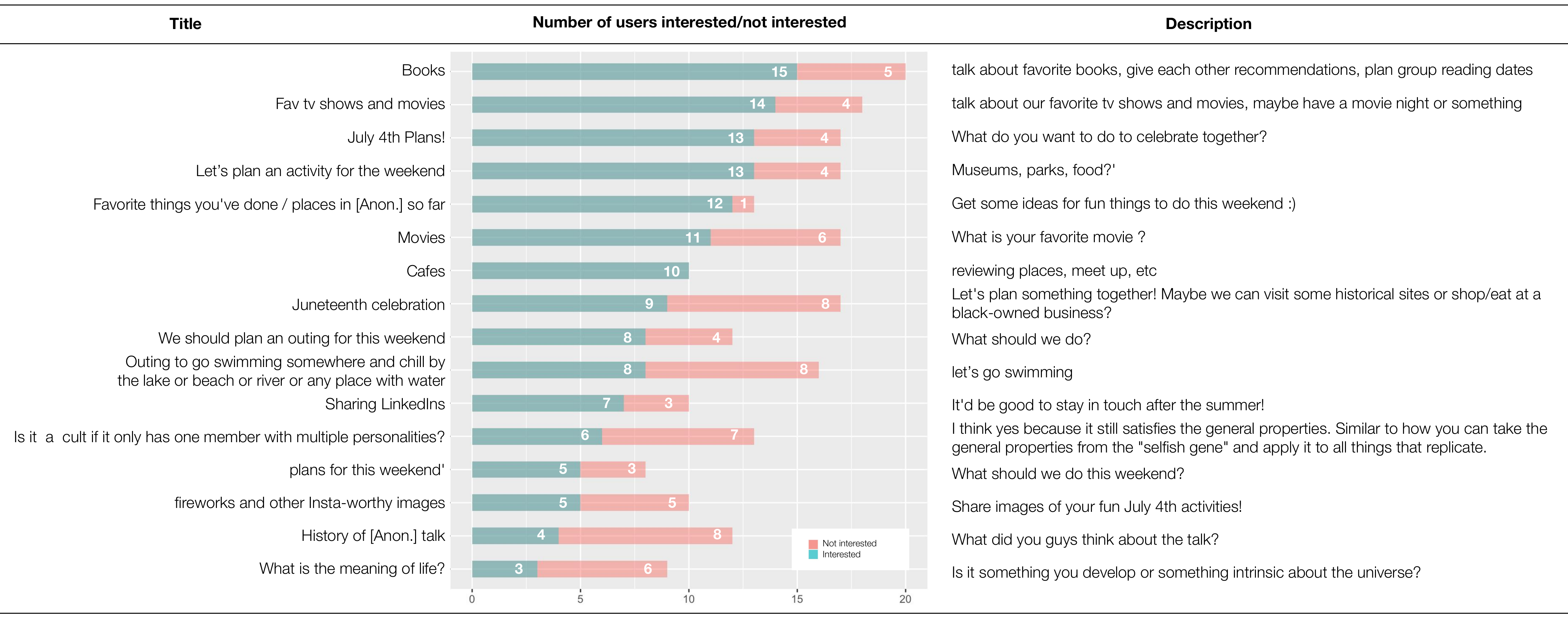}
    \caption{Nooks created by users. Of the $16$ nooks created by the participants, $10$ nooks were attempts at initiating offline activities. In $3$ of these $10$ nooks, the nook was created with a dual intention of facilitating interest-based interaction as well as planning a related activity. `fav tv shows and movies' was one such nook, described as a space to `talk about our favorite tv shows and movies, maybe have a movie night or something'. Each of these conversations had between $4$ and $15$ participants, with a median of $9$ participants. \ed{Here, the total number of responses against each nook indicates the number of participants that viewed and responded to the nook when it was incubating. Although every incubating nook was displayed on every users' \sys{} homepage, some nooks were viewed by fewer people (eg. `plans for this weekend' was viewed by $8$ people whereas 'Books' was viewed by $20$) because the frequency with which individuals viewed the homepage varied across individuals and with time.}}
    \Description{The figure shows a histogram with bars corresponding to each nook created by users of the system. The bars are arranged top to bottom in descending order of the number of members they had. From top to bottom, the nooks read: 1 “title:Books, description: talk about favorite books, give each other recommendations, plan group reading dates”. 2 “title:Fav tv shows and movies, description: talk about favorite books, give each other recommendations, plan group reading dates”. 3 “title:July 4th Plans!, description:What do you want to do to celebrate together?” 4 “title: Let’s plan an activity for the weekend, description: Museums, parks, food?” 5 “title: Favorite things you've done / places in [Anon.] so far, description: “Get some ideas for fun things to do this weekend :)” 6 “title: Movies, description: “What is your favorite movie ?” 7 “title: Cafes, description: reviewing places, meet up, etc ” 8 “title: Juneteenth celebration, description: Let's plan something together! Maybe we can visit some historical sites or shop/eat at a black-owned business?” 9 ”title: We should plan an outing for this weekend, description: What should we do? ” 10 “title: Outing to go swimming somewhere and chill by the lake or beach or river or any place with water, description: let’s go swimming ” 11 “title: Sharing LinkedIns, description: It'd be good to stay in touch after the summer! ” 12 “title: Is it  a  cult if it only has one member with multiple personalities?, description: I think yes because it still satisfies the general properties. Similar to how you can take the general properties from the ‘selfish gene’ and apply it to all things that replicate.” 13 “title: plans for this weekend, description:What should we do this weekend? ” 14 “title: fireworks and other Insta-worthy images, description:Share images of your fun July 4th activities! ” 15 “title: History of [Anon.] talk, description:What did you guys think about the talk? ” 16 “title: What is the meaning of life?, description:  Is it something you develop or something intrinsic about the universe?”}
    \label{fig:user-created-nooks}
\end{figure*}

Overwhelmingly, participants used \sys{} as a way to initiate offline activities (P2, P5, P6, P7, P8, P16, P17). Of the $16$ nooks created by the participants, $10$ nooks were attempts at initiating offline activities (Figure \ref{fig:user-created-nooks}). In $3$ of these $10$ nooks (Figure \ref{fig:user-created-nooks}), the nook was created with a dual intention of facilitating interest-based interaction as well as planning a related activity. For instance, one nook titled `fav tv shows and movies' was described as a space to `talk about our favorite tv shows and movies, maybe have a movie night or something'. Similarly, a nook titled `books' encouraged participants to `talk about favorite books, give each other recommendations, plan group reading dates'. In the remaining $7$ nooks, the intention was primarily to plan an activity. As an example, one nook was titled `Let's plan an activity for the weekend' that invited participants who were interested in visiting museums, park, or grabbing food. Nooks that focused on planning activities also generated more interest. The $10$ nooks that were about planning activities had an average of $11$ participants. In contrast, nooks that did not focus on planning activities had $7$ participants on average. 

Multiple participants described how they were primarily interested in connecting with others offline (P2, P5, P7, P8, P17) and that they used \sys{} to find new opportunities to do so. P8 mentioned how she was enthusiastic about \textit{``making things leave the group chat''} and so, was \textit{``more interested in nooks that could lead to something outside of the online context.''} Other participants mentioned similar motivations for creating and participating in nooks. 
\begin{quote}
Since there was like no one in my office---it was just me---I didn't get to interact with that many students in-person. So, I used [nooks] to make plans with them so that I could hang out with them later on. I thought it helped me a lot in talking more with others and hanging out with them which I know wouldn't have been possible if I relied on in-person [interactions], because, we're all kind of scattered across [campus].
\end{quote}

P17 mentioned how she was interested in hanging out with people and how \sys{} helped her meet `new people': \textit{"[\sys{} helped] with meeting new people in some ways. We both knew each other, because we have the weekly seminar in our program but we hadn't talked to each other, before.''} Similarly, P2 mentioned how a desire to be included in offline activities prompted her to participate in most \sys{}: \textit{``I am worried that people might plan stuff and then I'll miss out on it. I just join all the nooks to see what's happening''}.

However, not all participants were interested in the offline activities sparked by \sys{}, or even in connecting with others in the program. P20 mentioned how she didn't participate in a lot of nooks because her desires for social connection were fulfilled through alternate communities she was active in: \textit{``I do a lot of things so I meet people all the time, and so I kind of let my interest actually bring me to people like I went to salsa a lot. I was also living in a dorm with three people that I was getting closer to.''} As a result, she was picky about the nooks she chose to join. Another participant---P13---was local to the city and so, she already had friends in the city. Because of this, she didn't have as much time to participate in offline activities with others. She \textit{``wasn't really in [nooks] to make plans''.} Nevertheless, she participated in conversations about shared interests to learn more about others in the group: \textit{"just to see what I had in common with people and that's something that I bring up when I meet them in person.''}


\subsubsection{Strategies for using \sys{} to plan offline activities}
Participants repeatedly described how the collective practices that emerged around \sys{}, made it a uniquely effective means to \textit{``generate momentum''}(P16) for activities and events, and provided a \textit{``jumping off point to engage socially outside of slack''}(P6). P7 described how expressed interest within \sys{} increased chances of plans materializing: \textit{"[if I try to initiate plans in \#general], there's a chance I just get like an emoji reaction on [it] and I didn't want that, so I just used a nook because I knew that I would actually get a response from those [who] were actually interested''}. A nook provided a centralized venue to coordinate and make a plan. For instance, P2 mentioned:
\begin{quote}
It (my invitation) might get lost if I put it in general, so I was hoping that having a [activity-related] nook would be the kind of place where people would be centralized.
\end{quote}

The intentions established on joining a nook and the short duration of the conversations, accelerated scheduling: \textit{``because there was like that short time period, but also the nature of the conversation was about cafes to go to, the scheduling was something people had to figure out and did kind of make those plans"} (P16). 

Nooks were also used as a way to include more people in pre-existing plans. P17 noted how in one nook, \textit{"the person who created the nook said that she was going to some places with the student she works with and she was wondering whether other people were interested in joining them.''} Similarly P5 noted how a conversation with another participant led to the creation of a nook: \textit{"my project partner and I were thinking of playing tennis and then we were just like okay let's collectively create a tennis nook to see if people are interested.''}

Once an initial plan had gained momentum in a nook, participants also used alternate means to include more people. P16 mentioned how they manually added someone else to the nook because that they thought that the event might be of interest to them. In another case, the finalized plan was made public in the general channel in case others wanted to join: \textit{"with the fourth of July plan, we shared the chat inside general so we can get [more] people''} (P2). 

Finally, some participants mentioned how nooks often provided a starting point to create more persistent groups where plans were made on a recurring basis (P6, P8, P16). For instance P8 mentioned:
\begin{quote}
One nook that worked out well was a cafe nook and we actually did go out and get food together...a few people in the nook as well. Actually, the interest was started in the nook and we translated it into our food channel that's always open and just shared a broader like wider invitation for anyone else to come.    
\end{quote}

\subsubsection{Activity decreased as participants `found their people'}
Activity within \sys{} decreased after initial weeks. Participants attributed this to the fact that as they had a better understanding of each others' interests, and once they had formed closer relationships with some people, they didn't feel a need to use \sys{} (P5, P7, P8, P13). P5 mentioned:
\begin{quote}
I think for me personally, it was a really good initiation tool. Once we got to actually know some people I think I just didn't use [\sys{}] as much because I felt like I already knew who is going to respond to what. Later on, I could tell [who was interested in what].  
\end{quote}
P8 noted, \textit{``halfway through, a lot of people kind of found their little groups and have mostly stuck to them, myself included''}. When participants had already formed closer relationships, \sys{}' approach to facilitating conversations became constraining: \textit{``once you know your friends, you don't need to be able to text them between a 24 hour period and only about certain topics''}(P13). This is not necessarily a failure of \sys{}. As a tool to support relationship building between offline groups, \sys{} does not aim to sustain online interactions the way many social computing systems attempt to. In fact, participants accounts suggest that \sys{} had value as a tool for initiating interactions in groups.

\subsection{\sys{} provided new awareness about others and created opportunities to interact beyond \sys{}}
\label{sec:theme-4}
Conversations in \sys{} improved participants' understanding of each others' interests and often led to conversations beyond \sys{}. P13 mentioned how a conversation within \sys{} led to an extended conversation offline: \textit{``We were both in a nook about TV shows, and then we were just talking about it, the next day, and I was like oh, I saw your response in that as well, and so we were able to build off the conversation a little bit more''}. Similarly, P5 noted \textit{``I've even had conversations about [the nook topic], offline like not in a nook.''}

In many cases, participants were successful in initiating offline activities through conversations in \sys{} which led to new connections. P6 noted how a plan that emerged within a nook led to him meeting two new people. Similarly, P2 mentioned how she was able to get a group of people together to visit a park and watch fireworks. Participants mentioned how these experiences helped them get closer to \textit{``familiar strangers''} (P5, P7, P17). For instance, P5 noted, \textit{``these are people I had already met, but  in a very loose sense, you know. It's like I've seen them around but it's definitely not like we've been in really friendly situations before''}.

Beyond the awareness that emerged through participating in conversations, P5 mentioned how observing activity in \sys{} itself contributed an improved awareness of others' interests and their desire to socialize. P5 notes,\textit{"you don’t even need to interact with other people to learn what they are interested in. Because, you, swipe through the nooks and you see oh there's a group of people interested in like board games, or something. I think it's even useful just to gauge what the interests of your colleagues are''} (P5). \ed{None of the participants used the option to `send a message' to commonly encountered individuals within nooks. Instead, as our findings indicate, they used alternate pathways to continue interactions such as by scheduling offline activities through the nook or by approaching each other opportunistically offline. However, the signals from this feature still contributed to an improved awareness about colleagues' propensity for socialization}. By observing who was active in nooks\ed{---aided by the list of individuals they encountered most commonly---}participants were able to identify people that were interested in socializing: \textit{``It allowed me to see [who was] willing to connect with other people and would be down to maybe have a chat"} (P5).

\subsection{Deployment Limitations}
Our deployment revealed that participants were able to use \sys{} as a way to initiate online and offline interactions. That they personalized their use of the application to meet their needs is further evidence that they found it useful. However, our field deployment was not a field \textit{experiment}, because it did not include an appropriate control group. The lack of a control group prevents us from finding statistical benefits to using \sys{} (for example through a pre- and post-study comparison), because socialization outcomes are known to generally improve with participant tenure. As with other studies that trade off control for ecological validity, finding a control group in our setting is challenging, as social environments are complex and their evolution is path-dependent~\cite{salganik2006experimental}, such that dynamics among group members and their social relationships can yield widely varying states. 

\ed{Like other prior work  in sociotechnical systems~\cite{bernstein2010enhancing}, we also found it challenging to study non-usage. Although our interviewees had varying levels of usage (Figure \ref{fig:participant-stats-1}) and included participants that engaged minimally within \sys{} (P17, P20), participants that didn't join a nook at all (P23, P24, P25) did not participate in our interviews. As a result, though we were able to characterize some factors that led to reduced usage, we were unable to investigate why users might \textit{entirely} avoid interaction after signing-up on \sys{}. }

Some aspects of our deployment context also complicate the interpretation of our findings. Our participant pool was drawn among those participating in a summer research program in a private US-based research university, and does not cover the many possible ways that culture and training might have shaped their use of \sys{} and the benefits they accrued from it. \ed{The generalizability of our findings is also limited by the fact that we deployed \sys{} in an academic workplace. Academic workplaces such as research laboratories and academic departments have often been the sites of workplace studies in HCI research~\cite{plowman1995workplace}, representing sites of knowledge work~\cite{abdulgalimov2020designing, suchman1987plans, goodman1986collaboration, bellotti1993design}. We acknowledge that much like these prior efforts, findings from our deployment may not generalize or readily translate to design implications in non-academic work contexts. Academic workplaces tend to have weaker hierarchical power~\cite{musselin2007universities} and coworkers' tasks are more decoupled when compared to non-academic workplaces~\cite{musselin2007universities}. Additionally, when compared to individuals in non-academic workplaces, individuals in academic workplaces may identify more strongly with their disciplinary expertise (occupational commitment) than their institution (organizational commitment)~\cite{mcmurray2013determinants, teferra2014organized}. However, non-academic knowledge workplaces are evolving in ways that are increasing their similarity to academic workplaces. Hierarchies in non-academic knowledge workplaces are becoming flatter~\cite{rajan2006flattening} and as knowledge workers in non-academic workplaces desire greater autonomy, their occupational commitment is increasing at the expense of organizational commitment~\cite{may2002organizational}. Given these contemporary trends, we believe our findings can still offer some guidance when designing or deploying socialization tools in non-academic knowledge work contexts and the challenges we ran into will likely also surface in non-academic knowledge workplaces, but further investigation is necessary to tailor designs and deployment strategies to non-academic workplaces. Finally, the specifics of academic workplaces are not reflected in the \textit{design} of \sys{} and it can be integrated into any workplace that has a Slack workspace.}

\section{Discussion}
Through our deployment, we found that \sys{} was effective at supporting initial interactions between unacquainted individuals at work. In many cases, \sys{} promoted offline and online interactions outside the application. Here we reflect on open challenges and opportunities in the design of workplace socialization tools. 

\subsection{Tensions between supporting initial interactions and supporting interactions with ``insiders''}
Most of the participants in our deployment study were newcomers to the workplace. While \sys{} helped many participants in our study initiate new interactions and form new connections, the few participants in our deployment, who already had connections in the workplace, were less motivated to use \sys{}: \textit{``I had quite a few friends already in [workplace], so I wasn't as invested in these conversations''} (P13). This suggests that tools like \sys{} that support initial interactions, are more likely to attract newcomers and may especially support the formation of peer relationships among newcomer cohorts. Relationships with other newcomers are valuable---the peer support and collective sensemaking that these relationships enable, accelerate newcomer adjustment~\cite{zhou2022rookies}. Further, relationships with newcomers in different parts of the organization can provide access to distinct information and resources~\cite{zhou2022rookies, argote2000knowledge} while the resulting communication networks can support knowledge transfer within the organization. Peer-group bonds are commonly recognized as a necessary condition for successful organizational socialization~\cite{evan1963peer, zhou2022rookies}.

However, because those who already have connections across the organization do not readily see benefits of using tools like \sys{}, they may be less motivated to use them. Consequently, tools like \sys{} might be ill-suited to support interactions between newcomers and ``insiders''---those who have been in the organization for longer. Tensions caused by unequal distribution of benefits have been pointed out as a common challenge in designing groupware~\cite{grudin1994groupware} and \sys{} is unsurprisingly, prone to the same tensions. Cross-cohort relationships and relationships with ``insiders'', are important for newcomer socialization, however, supporting such relationships may require further institutional efforts.


\subsection{\sys{} is complementary to technology-mediated icebreaker experiences}
Researchers, designers, and artists have demonstrated the potential of technology-mediated icebreaker experiences to help relationship building in social groups such as summer camps, parties, and the workplace
~\cite{karahalios2004telemurals, hespanhol2014using, chlup2010breaking, depping2016trust, sileo1998strategies, trotto2013engage}.
Technology-mediated icebreaker experiences are typically one-off experiences that use online and offline approaches to bring people together and spark conversations by providing the group with a predefined conversation-starter~\cite{ludvigsen2005designing}. They provide this conversation-starter in various forms---discussion prompts~\cite{sileo1998strategies, chlup2010breaking}, installations~\cite{karahalios2004telemurals, snibbe2009social, hespanhol2014using}, gadgets~\cite{paulos2004familiar, mitchell2015sensing}, and even games~\cite{jarusriboonchai2016design, nasir2013cooperative}---and occasionally even sequence conversation-starters to spark increasingly intimate interactions~\cite{jarusriboonchai2016design}. Through this designed scaffolding, they contribute \textit{what} a group can talk about and create a comfortable atmosphere that supports progressive development of familiarity, solidarity and closeness~\cite{sprecher2013taking, jarusriboonchai2016design}. 

Unfortunately the same designed scaffolding that is intended to support initial interactions, can also limit the effectiveness of icebreaking experiences in some cases: predefined conversation-starters and sequences impede relationship building if some participants are forced into conversations they don't \textit{want} to have. Icebreaking experiences attempt to eliminate barriers around \textit{what} to talk about by enforcing predefined conversation-starters and sequences. However, even if participants are interested in the icebreaking experience as a whole, they may not be fully informed of the specific interactions that are to follow--- icebreaking experiences don't explicitly account for what each participant is willing to talk about, in a given group and at a certain point in time. As a result of this, icebreaking experiences can lead to situations where some participants in a conversation find the the topic uninteresting, irrelevant, or even inappropriate~\cite{miller2021meeting, rogers2002subtle}. Successful evolution of early relationships requires self-disclosures accompanied by reciprocal self-disclosures from others in the interaction~\cite{miller2021meeting, sprecher2013taking, svennevig2014direct}. However, if some participants find the conversation inappropriate or irrelevant, it can prevent them from sharing as well as reciprocating the self-disclosures they receive from others~\cite{miller2021meeting}. Additionally, when participants in an interaction fear that others might be uninterested and may not reciprocate their self-disclosures, they may choose not to participate themselves. 

Because they risk placing participants in conversations they find uncomfortable, designed icebreaker experiences---despite their benefits---can occasionally fail at their goal of relationship building. Prior work has argued for icebreaker experiences that are not over-prescriptive and that allow participants to appropriate and avoid interactions~\cite{mitchell2019levelling, rogers2002subtle}. \sys{} takes this complementary approach by instead allowing individuals to define for themselves what they want to talk about with new acquaintances.

\subsection{Designing and communicating new organizational routines}
Systems for groups (groupware), invariably attempt to alter group practices and as a result, face resistance. Prior work has systematized factors that influence the acceptance of groupware, recognizing that the success of groupware is sensitive to how it is introduced in groups and how it challenges their current practices~\cite{grudin1994groupware, ehrlich1987strategies}. Most of this work emphasizes strategies to shape practices that are enacted individually--- they attempt to help each individual adapt to a new tool. Similar to most groupware, finding contextually appropriate strategies to communicate new individual practices are critical to the successful deployment of \sys{}. However, \sys{} requires acceptance along an additional dimension: it also attempts to introduce a repetitive and collectively enacted practice. Because \sys{} is a small scale social computing system, we enforced a time-driven schedule as a way to lower interruption costs in pushing the sparse interaction opportunities to users~\cite{10.7551/mitpress/8472.003.0007}: incubating nooks were made available at 4pm and nooks were activated at 12pm. For \sys{} to be effective, it required users across the workplace to adapt to a new collective practice---all users needed to express their interest in incubating nooks between 4pm and 12pm. This introduced a new routine in the workplace---it required \textit{groups} as a whole to enact repetitive and interdependent actions~\cite{gersick1990habitual}. While the specific routine may vary, any tool that attempts to orchestrate periodic interactions within a group, has to also design and communicate a new organizational routine. To increase the acceptance of a new routine, we made decisions about when to activate nooks and when to notify users about incubating nooks based on early observations about distributions of users' work days. These decisions can be highly context specific. Further, the groups that \sys{} was tested and deployed in, all involved individuals in the same timezone. If teams are distributed across timezones, and as a result workdays diverge significantly, introducing a similar routine can be challenging because there may be no `right' decisions. We communicated the new routine by highlighting it in the seminar. Additionally, we injected the system with predefined nooks as a way to facilitate familiarization with the new routine even if users were not contributing nooks themselves. In many ways\ed{,} our approaches to communicating the new routine extend strategies devised earlier in the context of groupware, specifically the use of educational material and training~\cite{ehrlich1987strategies}, to the new task of communicating routines.
\ed{\subsection{Would \sys{} improve organizational effectiveness?}

\sys{} was aimed at supporting social relationships at work, which is known to improve many conditions---such as motivation~\cite{kanfer1990motivation}, resilience~\cite{dutton2003power}, and learning~\cite{lave1991situated}---under which organizations are effective. Our deployment study suggests \sys{} has the potential to support relationship building at work and therefore, could improve organizational effectiveness. However, it might seem that \sys{}, much like email, IM, and other communication technologies, also has the potential to undermine organizational effectiveness if it leads to excessive interruptions that are disruptive and cause loss of focus. Would \sys{} actually improve organizational effectiveness? We can't conclusively answer this question, as we did not measure the impact of \sys{} on metrics of organizational effectiveness, such as productivity and focus. This was in part because it is challenging to reliably measure stable changes to organizational effectiveness in a $9$ week deployment study, as there is often a time lag between organizational change and realization of impacts~\cite{davern2000discovering, miller1977improve}. A long-term deployment might be more apt to measure impacts of \sys{} on organizational effectiveness. 

Even though our current study did not generate evidence about the effects of \sys{} on measures of effectiveness (such as productivity or focus), we believe the potential negative impacts of \sys{} on organizational effectiveness, are limited for three reasons. First, by incubating nooks in batches, the design of \sys{} attempts to minimize interruptions by \textit{batching} notifications---a strategy that can limit the harms of information interruptions on workplace productivity~\cite{mark2016email}. Second, \sys{} recognizes that patterns of attention and focus may vary throughout a day and across individuals~\cite{mark2018effects, hudson2002d} and so, instead of identifying opportune moments to orchestrate synchronous conversations, \sys{} supports asynchronous conversations that individuals can participate in whenever they experience a lull during the day, lowering disruption to their other work. Finally, evidence suggests that people adapt to new technologies and sources of distraction at work, without the need to eliminate them~\cite{rouncefield1994working, webster1997audience, mark2018effects}. Over time, individuals learn to self-regulate their practices~\cite{rouncefield1994working} and technology use~\cite{laumer2017challenge} to maintain face-to-face and technological distractions at their desired level. Because, \sys{} does not force use in any way and allows individuals to use it to the extent they desire, individuals can adapt their use of \sys{} over time so that its negative impacts on their work are low. 
}
\subsection{Future work}

\subsubsection{Matching peers over interests as a way to promote heterophilous interactions at work}
Social interactions are often driven by homophily; people tend to gravitate towards others within their demographic groups. At work, for instance, people are more likely to seek others of similar ethnicity and gender~\cite{bacharach2005diversity}. This can often cause exclusion in the workplace, leaving minorities feeling even more isolated in the workplace~\cite{karimi2018homophily}. \sys{} demonstrates the promise of revealing shared interests in online spaces as a way to promote inclusive conversations by making group boundaries less salient. By shifting focus away from \textit{who} is part of the conversation and towards \textit{what} unites them, \sys{} has the potential to support workplace interactions between demographically dissimilar individuals. Orchestrating interactions between demographically diverse individuals is especially useful when they are new to the organization because it is these early interactions within an organization that lead to homophilic networks to begin with~\cite{karimi2018homophily}. Future work can explore the design of tools that build on \sys{} and have particularly powerful benefits for fostering peer relationships among demographically diverse individuals at work.

\subsubsection{Facilitating socialization among newcomers and `insiders'}
\sys{} was effective at promoting initial interactions at work but only among those that were looking for new connections. \ed{As a result,
Nooks may fall short when it comes to forging connections between newcomers and ‘insiders’ in an organization. However, it may be possible to motivate use by insiders and forge relationships between newcomers and insiders through: (1) organizational communication campaigns; and (2) intentionally crafted practices. Informed by prior work, we present some approaches along both of these dimensions that may increase participation from insiders, and that we hope to explore further:
\begin{enumerate}
    \item \textit{Designing communication campaigns that motivate use by insiders:} Communication strategies can be effective at clarifying the purpose of workplace technologies, updating employees' beliefs, and motivating use~\cite{louw2013guiding, bin2010removing}. Prior work has emphasized the importance of leadership buy-in for acceptance of social technologies at work~\cite{laumer2017challenge, dimicco2008motivations,bin2010removing} and so, organization-wide communication campaigns that make endorsement by leadership visible to the employees can increase acceptance and use of social technologies among insiders. When benefits are unevenly distributed---benefits to newcomers are larger than benefits to insiders---drawing attention to the \textit{collective} benefits can also motivate all users~\cite{grudin1994groupware}. To motivate insiders to use \sys{}, communication campaigns can emphasize how \sys{} can benefit the organization as a whole by facilitating denser interactions in the workplace community, and paving the way for knowledge sharing and collaboration.  
    \item \textit{Crafting practices that routinely engage insiders:} Another approach to increase participation from insiders is introducing new organization-wide practices that routinely draw insiders to \sys{}. Crafted practices, and especially those that are playful, can be powerful for overcoming inertia and driving use of social technologies in the workplace~\cite{laumer2017challenge}. To promote use of its enterprise social network (ESN), IBM implemented a `go viral' campaign which included a weekly playful activity~\cite{laumer2017challenge}. This stimulated ESN use because employees, in addition to enjoying participation, were also keen to involve their co-workers~\cite{laumer2017challenge}. Similarly, one way to increase acceptance and use among insiders could be to designate \sys{} as the default space to host conversations about co-occuring community-wide events or experiences~\cite{cong2021collective} (e.g. a guest talk or a hackathon). Making it a venue for such conversations can attract older employees to \sys{} who may be interested in talking about community experiences. In this case, even if forming new connections is not insiders' primary motive to interact, interactions about community experiences can promote feelings of connectedness and create opportunities for newcomers to connect with insiders. As another practice, the application could nudge the few insiders that already interact within nooks, to pull their close connections who are inactive within nooks, into conversations that might be of interest to them. Insiders would be more familiar with the interests of their close connections and might be better able to identify when they might appreciate being included.     
\end{enumerate}

These dimensions are complementary and often interdependent. For example, a communication campaign that encourages insiders to include their inactive colleagues in nooks, could help the practice gain momentum. Finally, to design solutions that are aligned with organizational values and that organizations are willing to deploy in practice, future work can involve relevant organizational stakeholders, adopting co-design~\cite{sanders2002user} and value-sensitive design~\cite{zhu2018value} approaches to imagine solutions along both dimensions.    
}


\ed{\subsubsection{Scope for personalization}
In our current implementation of \sys{}, incubating nooks are ordered by the time at which they are created. Instead, if we incorporate a model of a user's interests, we can rank the incubating nooks so that the most relevant incubating nooks are shown to the user first. Users' expression of interest and non-interest in nooks can provide a feedback signal to update models of user preferences. More generally, future work can explore how a recommendation system, driven by a model of user preferences, can lower the effort of identifying interesting nooks in an incubating batch. This might be particularly beneficial as the size of the workplace and the number of nooks scale.}




\section{Conclusion}
In this paper, we introduce \sys{}, an online tool to support initial interactions in the workplace by enabling individuals to discover and interact with others they share common interests with, while lowering risks of social evaluation. By convening conversations over shared interests, on an ongoing basis, \sys{} supports relationship building in the workplace. A nine-week field deployment demonstrated that \sys{} can catalyze new conversations both online and offline, and provide an ambient awareness of others' interests and desires for socialization. Additionally, participants personalized their use of \sys{} to meet their own needs. Based on our findings and our experience developing and deploying \sys{}, we discussed challenges and future directions for the design of workplace socialization tools. 

\begin{acks}
We are grateful to Laura Dabbish and Diana Rotondo for their support in deploying \sys{} with the REU program at the Human-Computer Interaction Institute, Carnegie Mellon University, and the many students in the REU program who used Nooks to connect with their peers. We would also like to thank Kimi Wenzel, Yasmine Kotturi, and Julia Cambre for their feedback on earlier drafts of this paper. This work was supported by the Office of Naval Research but this article solely reflects the opinions and conclusions of the
authors and not our funder.
\end{acks}

\bibliographystyle{ACM-Reference-Format}
\bibliography{bibliography}

\end{document}